\documentclass[conference]{IEEEtran}
\IEEEoverridecommandlockouts
\usepackage{cite}
\usepackage{amsmath,amssymb,amsfonts}
\usepackage{algorithmic}
\usepackage{graphicx}
\usepackage{textcomp}
\usepackage{xcolor}
\usepackage{float}
\def\BibTeX{{\rm B\kern-.05em{\sc i\kern-.025em b}\kern-.08em
    T\kern-.1667em\lower.7ex\hbox{E}\kern-.125emX}}

\begin{document}
\title{EmoSens: Emotion Recognition based on Sensor data analysis using LightGBM\\
}
\makeatletter
\newcommand{\linebreakand}{%
  \end{@IEEEauthorhalign}
  \hfill\mbox{}\par
  \mbox{}\hfill\begin{@IEEEauthorhalign}
}
\makeatother

\renewcommand\IEEEkeywordsname{Keywords}

\author{
\IEEEauthorblockN{Gayathri S\IEEEauthorrefmark{1},
Akshat Anand\IEEEauthorrefmark{2}, 
Astha Vijayvargiya\IEEEauthorrefmark{3},
Pushpalatha M\IEEEauthorrefmark{4} and 
Vaishnavi Moorthy\IEEEauthorrefmark{5}}
\IEEEauthorblockA{School of Computing,
SRM Institute of Science and Technology,\\
Kattankulathur,Tamil Nadu– 603203,India \\
Email: \IEEEauthorrefmark{1}sz5928@srmist.edu.in,
\IEEEauthorrefmark{2}aa4407@srmist.edu.in,
\IEEEauthorrefmark{3}av4681@srmist.edu.in,
\IEEEauthorrefmark{4}pushpalm@srmist.edu.in,
\IEEEauthorrefmark{5}vaishnavim@srmist.edu.in}
\and
\linebreakand 
\IEEEauthorblockN{Sumit Kumar\IEEEauthorrefmark{6},
Harichandana BSS\IEEEauthorrefmark{7},
\IEEEauthorblockA{Samsung Research Institute,
Bengaluru,India \\
Email: \IEEEauthorrefmark{6}sumit.kr@samsung.com,
\IEEEauthorrefmark{7}hari.ss@samsung.com}}}

\IEEEoverridecommandlockouts
\IEEEpubid{\makebox[\columnwidth]{978-1-6654-9781-7/22/\$31.00~\copyright~2022 IEEE \hfill} \hspace{\columnsep}\makebox[\columnwidth]{ }}
\maketitle
\IEEEpubidadjcol

\begin{abstract}
Smart wearables have played an integral part in our day to day life.From recording ECG signals to analysing body fat composition,the smart wearables can do it all. The smart devices encompass various sensors which can be employed to derive meaningful information regarding the user’s physical and psychological conditions.Our approach focuses on employing such sensors to identify and obtain the variations in the mood of a user at a given instance through the use of supervised machine learning techniques.The study examines the performance of various supervised learning models such as Decision Trees, Random Forests, XGBoost, LightGBM on the dataset. With our proposed model, we obtained a high recognition rate of 92.5\% using XGBoost and LightGBM for 9 different emotion classes.By utilizing this, we aim to improvise and suggest methods to aid emotion recognition for better mental health analysis and mood monitoring.

\end{abstract}


\section{Introduction}
 Affective computing involves studying and developing intelligent systems that can identify, understand and imitate human emotions. It is a multidisciplinary field that spans across  computer science, cognitive science and psychology. The field of affective computing was first discussed in the year 1995 in Rosalind Picard’s work \cite{b1} which elaborated on the ability of computers to imitate human emotions by understanding the underlying expression. It is necessary to study the emotions of humans as they hold a huge significance in human interactions as suggested by Ekman \cite{b2}. The fact that humans treat computers as other living creatures necessitates improving human-computer interaction. \cite{b3}. This solidifies the idea that the interactions between humans and computers can be made naturalistic through the use of affective computing. In other words, affective computing helps in instilling human essence in artificially intelligent machines. Research in affective computing can be categorized into two: sentiment analysis and emotion recognition.Our objective is to focus on emotion recognition. 
 
 Emotion recognition has always been a thriving topic for areas of research as it deals with human interaction which in fact helps understanding human emotional states better with use of engineering, psychology and cognitive science. Recognizing emotions through these modalities could be helpful in accurately understanding psychological health and humans with computer interaction without using external medical instruments. Emotion recognition has numerous applications. It can help monitor emotional states of humans in critical situations. Under clinical situations, it can be used to monitor psychological conditions of patients.In the entertainment/ video game industry, it can be used to recognize the emotions of users to a particular video, film clip or game. In the consumer service industry to improve marketing or enhance user experience by identifying the user’s reaction to a product. Since these applications of emotion recognition are highly user-centered, it is only fair if the solution is easily accessible, low cost and computationally efficient. The proposed approach aims to integrate the vast availability of smart edge devices and advanced machine learning techniques to make emotion recognition at a user-centered level feasible.
 
 Emotions are complex and encompass changes in feelings, thoughts, behaviour, poise, body language, cognitive reactions. These changes can be recognized through various audio and visual cues. Most studies use these cues to identify the emotional states of a human at a given situation. However, facial expression and tonal variations can be subject to manipulation as humans tend to control emotions. Therefore, a better and more accurate indicator of emotional state would be physiological signals. The physiological signals are directly related to the autonomous nervous system (ANS) which is a part of the peripheral nervous system of the human body. The ANS sympathetic nerves are triggered when a human goes through a positive or negative experience. This affects the heart rate significantly. There are a number of physiological signals that are used to perform emotion recognition.  A few popular ones include ECG(Electrocardiogram), EEG(Electroencephalogram), GSR(Galvanic Skin Response), EMG(Electromyography), HRV(Heart Rate Variability). Out of these, the ECG is most popularly used as it indicates emotions, less susceptible to noise and relatively easy to access.
 
 Everything a person feels or does has an immediate impact on their heart.ECG is essentially obtained by studying the continuous contractions and expanding of the heart. Hence, the heart rate conveys the variations in emotions of a person. The sympathetic nervous system is triggered due to variability in emotion which in turn enhances perception to internal and external stimuli. These changes can be detected in ECG. Apart from being indicative of the variations in emotions of a human, heart rate can be easily measured using the sensors in modern smart watches. Not only would this ensure a user-centered approach towards emotion recognition and monitoring, but is also cost-efficient. However, the acquisition of data and processing it for further use is not as easy it sounds as the signals and heart rates are subject to variations within the same subject (intra-subject) and between different subjects(inter-subject), apart from the noise due to change in posture, monitoring environment and devices used to record the signals\cite{b31}. 

With the complexity of data acquisition and processing being high, most datasets used for the task of emotion recognition have similar methods of data acquisition. A common observation is that most of the datasets collect data from a small number of volunteers who are stimulated using emotion inducing audio/video clips and the subsequent reaction in the required physiological signal signal  that follows is recorded through an appropriate sensor. The emotion labelling is performed through Self Assessment Manikins (SAM) \cite{b4}. Very commonly used datasets for the purpose of emotion recognition are DREAMER \cite{b5}, MEPD \cite{b6}, MAHNOB-HCI \cite{b7}, K-Emocon \cite{b8}, AMIGOS \cite{b9}. The proposed work was performed with the DREAMER dataset. Appropriate pre-processing, feature extraction, training and validation methods were observed and performed on the DREAMER dataset to achieve the required results.
The study contributes the following: a user-centered approach to develop an emotion recognition system using
\begin{enumerate}
\item Low-cost, non-invasive, wearable sensors 
\item Classifying emotions using machine learning 
\end{enumerate}
We have explained in detail about the workflow for developing our emotion recognition system in the upcoming sections. The classification approaches that were considered include K-Nearest Neighbours, Linear SVM, RBF SVM,  Decision Trees, Random Forests, XGBoost, LightGBM, AdaBoost.

The structure of the following sections are as follows: Section II contains the review of prior arts relevant to our proposed work. Section III goes over the dataset,the preprocessing techniques used. Section IV discusses emotion models and the target variable. Section V presents the classification models and their results. Section VI concludes the study and proposes future work.

\section{Literature Review}
Various physiological modalities are used across multiple studies for emotion recognition.  ECG has been widely used to monitor and assess psychological and mental conditions \cite{b10}. It can act as a powerful indicator of stress in the human body \cite{b11}\cite{b12}. As a result,the proposed work is interested in ECG-based emotion recognition, we only considered the ECG modality of the available datasets. Among the available datasets we considered AMIGOS, K-Emocon, MAHNOB-HCI, MEPD, DREAMER.

After examining the above datasets and the features, we concluded that the DREAMER dataset fit our needs the best.

Ekman suggested that emotions are discrete and measurable  \cite{b13}. On that basis he summarized emotions into 6 elementary emotions including happy, anger, sad, fear, disgust and surprise  \cite{b14}. This proposed model is called the Discrete Emotion Model.The proposed study aims to identify discrete emotions through ECG signals. A study by Dissanayake et al. \cite{b15} recognized 6 emotional classes through HRV features obtained from ECG signal. An accuracy of 80\%  was achieved using Extra Tree Classifier with feature selection. KNN gave a 52\% accuracy on a study by Jerrita et al. \cite{b16} on 30 subjects. Video clips were used for emotion elicitation. 5 class emotional classification on 25 subjects achieved an accuracy of 56.9\% using SVM by Guo H. at al.\cite{b17}. A study conducted by Zhang et al.  \cite{b18} with 4 emotional classes into consideration provided an accuracy of 92\%. The best results were provided by a KNN model after feature extraction. 6 classes of emotions were classified with an accuracy of 92.87\% by Selvaraj et al.  \cite{b19}. Only non-linear features of ECG signal were considered. Fuzzy KNN performed the best. Study by Yang et al.  \cite{b20} utilizes a Bayesian Network on ECG and EEG multimodal sensor data to classify 6 emotion classes with an accuracy of 98.06\%. 

We propose an approach in Fig.~\ref{fig:modelarch} that follows the traditional machine learning pipeline to improve the emotion recognition performance of 9 emotion classes instead of the 6 emotion classes that are considered in most of the prior works. We have achieved a benchmark accuracy of 92.5\% for 9 emotion classes using our proposed approach. Our ECG signal based approach results are reported against the publicly available dataset DREAMER. The signals are acquired using low-cost, wearable sensors which represent the dependability of our proposed methodology in a real-world environment. 

\begin{table}[htbp]
\footnotesize
\caption{Comparison with prior arts}
\begin{center}
\begin{tabular}{|l|l|l|l|}
\hline
\cline{2-4} 
\textbf{\textit{Paper}}& \textbf{\textit{Model}}& \textbf{\textit{Classes}}& \textbf{\textit{Accuracy(\%)}} \\
\hline
Yang et al. \cite{b20}&  Bayesian Network& 6& 98.6\\
\hline
Zhang et al. \cite{b18}&  KNN& 4& 92\\
\hline
Selvaraj et al.  \cite{b19}&   KNN& 6& 92.87\\
\hline
Dissanayake et al. \cite{b15}&   Extra Tree Classifier& 6& 80\\
\hline
Jerrita et al.  \cite{b16}&   KNN& 6& 52\\
\hline
Guo H et al. \cite{b17}&   SVM& 5& 56.9\\
\hline
\textbf{\textbf{Proposed Technique}}&   \textbf{\textbf{LightGBM}}& \textbf{\textbf{9}}& \textbf{\textbf{92.5}}\\
\hline
\end{tabular}
\end{center}
\end{table}

\begin{figure}[htbp]
\centerline{\includegraphics[scale=0.2889]{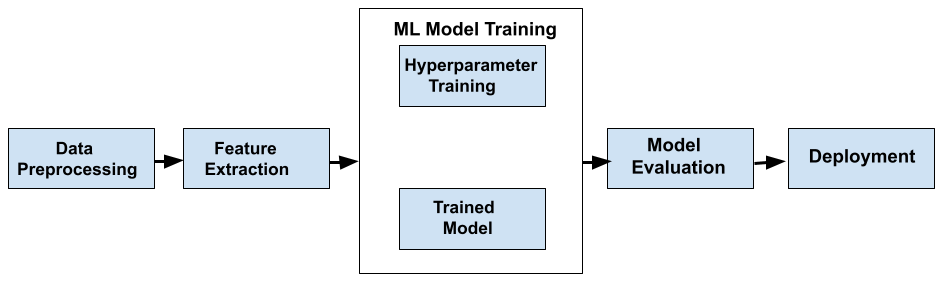}}
\caption{Proposed model}
\label{fig:modelarch}
\end{figure}

\section{Dataset construction and features}
We looked at several datasets such as DREAMER, MPED, K-emocon out of which we found DREAMER as the perfect dataset to fit our needs. The features which attracted us to go with DREAMER dataset were
\begin{enumerate}
    \item The dataset focused on extensive affect recognition.
    \item It uses low cost, off-the-shelf devices to measure ECG and EEG signals.
    \item The results using the devices was on-par with medical grade devices in terms of affect(emotion) recognition.
    \item It aims to integrate emotion recognition with daily life activities which was very well aligned with our research. 
\end{enumerate}

\subsection{Data acquisition}
The DREAMER dataset contains EEG and ECG signal data from 23 volunteers out of which 14 are male and 9 are female.The volunteers were aged between 22-33 years. The data was collected through audio and video stimuli. The study was conducted in a darkened room with a 45” monitor to play the audio and video stimuli. 18 film clips suggested by Gabert-quillen et al. \cite{b29} that evoke different emotions from the spectator were played for 65-353 seconds, out of which reactions from the last 60 secs were taken into consideration to accommodate the change in emotions a person goes through. The 18 clips targeted 9 emotions, namely: amusement, excitement, happiness, calmness, anger, disgust, fear, sadness and surprise. Out of the EEG and ECG data obtained, we have chosen to focus on ECG signals. These features will be discussed in the upcoming sections. 

\subsection{ECG signal and features}
A SHIMMER wireless sensor was used to record the ECG signals at 256 Hz. It is shown in studies that ECG correlates with the emotional state of a human \cite{b21}\cite{b22}. The Heart rate and heart rate variability have been noted to have an association with the emotional state of a person. The spectral, temporal and frequency parameters of HRV are useful in detecting emotions. These features were derived using Pan-Tompkins QRS detection algorithm \cite{b28}. The algorithm is used to detect the QRS complexes present in the ECG signal. These complexes can further be used to detect the R peaks. The Augsburg Biosignal Toolbox \cite{b23} was used to get the mean,median , standard deviation from the PQRST complexes. The BioSig Toolbox \cite{b24} was used to obtain the HRV features for further analysis. These features included RR intervals, RMSSD, pSD, LF,HF, LF/HF. The authors extracted 71 features from the ECG signal. Further,baseline normalization was performed to ensure well distributed data points.

\begin{figure}[htbp]
\centerline{\includegraphics[scale=0.18]{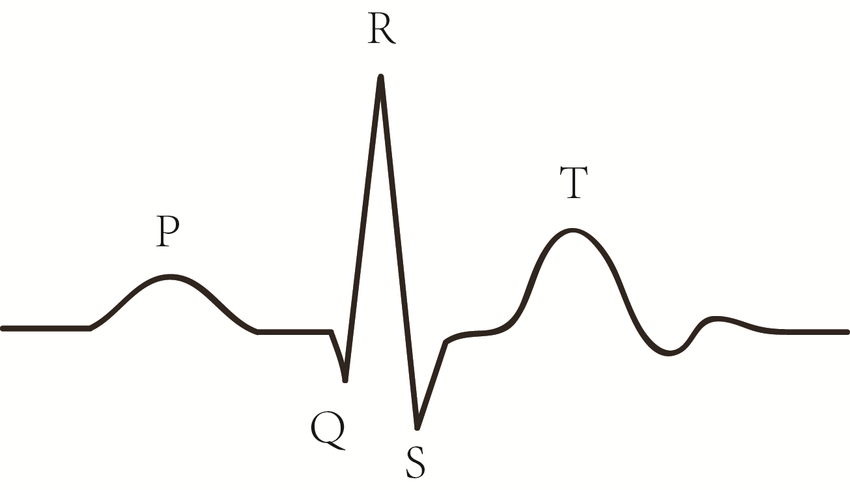}}
\caption{PQRST Complex for ECG \cite{b30}}
\end{figure}

\subsection{Data preprocessing and feature extraction}
The dataset was made in the MATLAB environment, so it was in .mat format. In order to process it we use the Neurokit library. We converted the .mat file into .csv by loading all Neurokit dependencies. After conversion we got 34 features based on Heart Rate Variability. After thorough exploratory data analysis we found that video\_name wasn't an important feature and we decided to drop it. 

\begin{figure}[htbp]
\centerline{\includegraphics[scale=0.3]{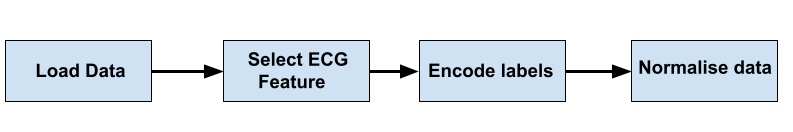}}
\caption{Data Pre-processing Workflow}
\end{figure}

\section{Emotion models}
There are two types of emotional models: Discrete Emotional Model and Dimensional Emotional Model. The Discrete Emotional Model assumes that human beings go through a set of basic emotions that are recognizable through expressions and biological processes. These emotions are considered to be a category rather than a stand-alone emotional state. In the Discrete Emotional Model(DEM), emotions are categorized as happiness,fear, anger, disgust, sadness and surprise. 

The Dimensional Emotional Model preconceives emotions experienced by humans by describing them in two dimensional or three dimensional planes. The model also uses valence and arousal to indicate the intensity of the emotion. It suggests that emotional states of humans are affected by a complex, intertwined neurophysiological network.There are various dimensional models suggested. A few popular ones include the Circumplex model\cite{b25}, Vector model, Plutchik’s model \cite{b26}, Positive activation- negative activation model \cite{b27}. 

\begin{figure}[htbp]
\centerline{\includegraphics[scale=0.29876]{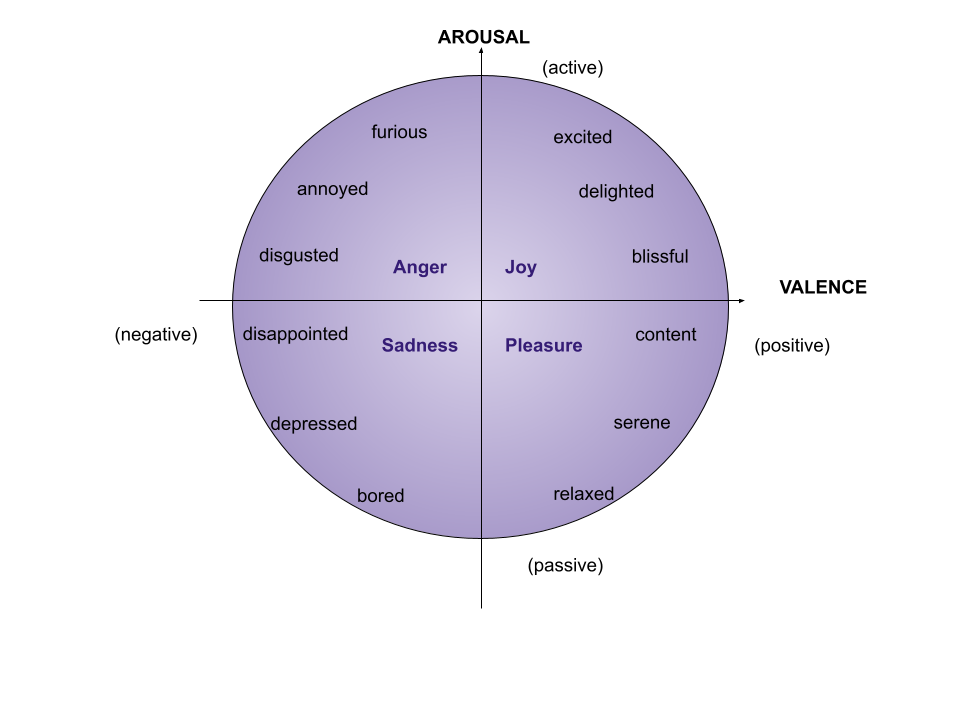}}
\caption{Circumplex model of emotions}
\end{figure}
The dataset also contains valence and arousal values which are provided emphasis in the dimensional model. The dataset gathers valence and arousal scores on a scale 1 - 5,pertaining to the intensity of the emotion experienced by the volunteer. Valence represents how positive or negative the reaction to the stimuli was. Arousal represents how calm or excited a human is. A valence score of 1 (low valence) indicates a negative reaction and a score of 5 (high valence) indicates a positive reaction. An arousal score of 1 (low arousal) indicates calmness reaction and a score of 5 (high arousal) indicates excitement. 

\subsection{Target variable}
Our target variable was the target emotion feature. The task is to classify the emotions into 9 different categories. The emotions under consideration are: Calmness, Surprise, Amusement, Fear, Excitement, Disgust, Happiness, Anger, Sadness. The classes are balanced. Our approach follows the discrete emotional model.

\section{Model description \& hyperparameter tuning}
We used various machine learning models to categorize the emotions according to the ECG signal readings. The models that we used were: K-Nearest Neighbours, Linear SVM, RBF SVM,  Decision Trees, Random Forests, XGBoost, LightGBM, AdaBoost. Apart from this we also used a Multilayer Perceptron Classifier with an adaptive learning rate. 

\subsection{Model training}
For training and validation, we have used GroupKfold validation in Scikit learn to ensure that overfitting is combated. GroupKFold ensures that the number of distinct groups is the same in each fold, thereby ensuring class balance. We have used 10 splits and the classification metrics were averaged for all of them at the end of the training procedure. A pipeline was created which first used min-max scaling on the dataset for normalization and then performed classification on the dataset using 11 different classifiers. Following this, the mean accuracy score,precision, recall, F-score and runtime were calculated for the 10 different splits of the dataset.

We performed hyperparameter tuning on all the machine learning models and observed that the accuracy increased for Random Forest, Decision Trees. The updated results of these can be found in Table III.

\section{Results and Inferences}
The proposed approach calculated the accuracy of multi emotion classification to evaluate the performance of these classifiers. We used 34 HRV features of the dataset. GroupKFold cross-validation was used and n\_splits was set to be 10. The average accuracy for these 10 splits were then identified and tabulated in Table II. Further to improve the accuracy of the classifiers, we performed hyperparameter tuning on them and noticed that there was an improvement in performance of Decision Trees and Random Forests. The comparison of the accuracy for the hyperparameter tuned models is given in Table III. Out of all the classifiers, XGBoost and LightGBM performed well to give an accuracy score of 92.5\%. Decision Tree gave an accuracy of 92.5\% after hyperparameter tuning. Random Forests gave an accuracy score of 92.2\%. This signifies that the task of emotion recognition and classification can be performed efficiently using advanced machine learning classifiers such as XGBoost, LightGBM, Decision tree or Random Forests. All the 4 classifiers gave comparable results with a small average runtime.The training curves for these graphs indicate how the model’s accuracy progressively increases for the validation set. Support Vector Machines (SVM), which was a popular choice in prior arts, did not give the expected results.In Fig. 5. we can see the learning curve for SVM multiclass classifier.A reason for the low accuracy could be the number of classes.From Fig. 9. we can observe that RBF SVM performs better than Linear SVM which is a clear indicator that the data is not linearly separable.This also implicitly addresses the overall underperformance of SVM for our proposed approach.
Fig. 6. indicates how decision tree performs on the training and validation sets. It can be seen that even though the model overfits on the training data, the validation accuracy seems to be increasing with the increasing number of training iterations. It shows that there is some generalization happening. Similarly, LightGBM in Fig. 8.performs well on the validation set to give an accuracy of 92.5\%. LightGBM is a gradient boosting framework based on decision trees which is fast and gives improved accuracy. It can handle large amounts of data, multiclass classification and requires low memory. 

\begin{figure}[htbp]
\centerline{\includegraphics[scale=0.7]{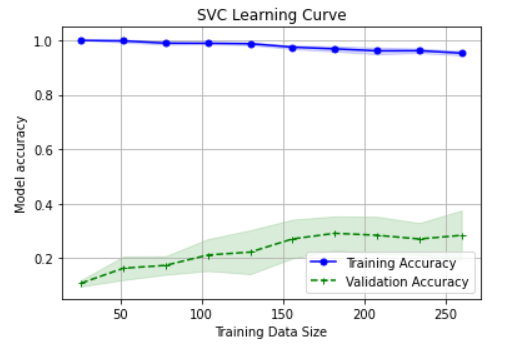}}
\caption{SVM training curve}
\end{figure}

\begin{figure}[htbp]
\centerline{\includegraphics[scale=0.7]{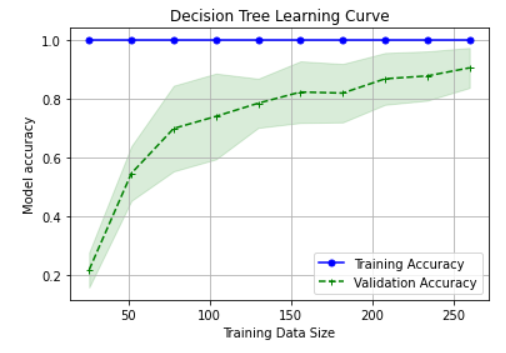}}
\caption{Decision Tree training curve}
\end{figure}

\begin{figure}[htbp]
\centerline{\includegraphics[scale=0.7]{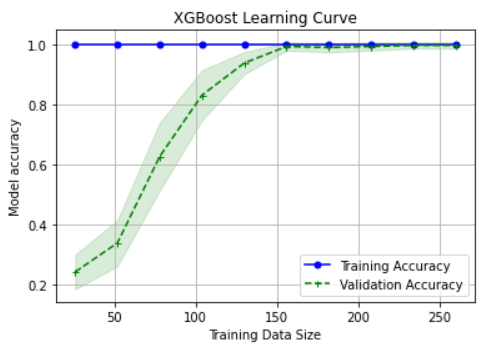}}
\caption{XGBoost Training curve}
\end{figure}

\begin{figure}[htbp]
\centerline{\includegraphics[scale=0.7]{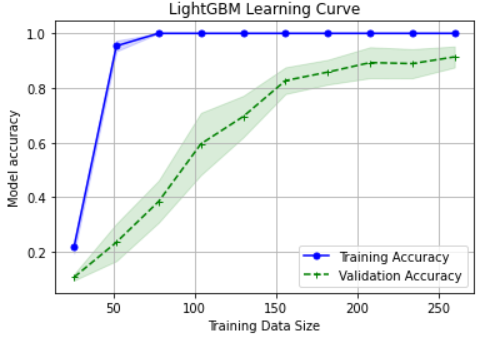}}
\caption{LightGBM Training Curve}
\end{figure}

\begin{table}[htbp]
\caption{Performance of classifiers}
\begin{center}
\begin{tabular}{|c|c|c|c|c|}
\hline
\cline{2-4} 
\textbf{\textit{Model}} & \textbf{\textit{Accuracy (\%)}}& \textbf{\textit{Precision}}& \textbf{\textit{Recall}}& \textbf{\textit{F-score}}\\
\hline
Nearest Neighbours& 43.33& 46.17& 43.3& 40.16 \\
\hline
Linear SVM&   	26.67&      22.64&  	26.67&    21.38\\
\hline
RBF SVM&   	55.28&   65.84&	  55.28&        54.19\\
\hline
Gaussian Process&    36.11&      28.04&  	36.11&  	30.79\\
\hline
Decision Tree&   	87.50&     88.5&  	87.5&     87.14\\
\hline
Random Forest&   	91.39&  	91.00&      91.39&   	91.11\\
\hline
Neural Net&  	41.11&  	41.1&      41.11&   	36.28\\
\hline
AdaBoost&   	22.2&      11.57&  	22.22&   	10.85\\
\hline
Naive Bayes&   	25.0&      20.99&  	25.0&     19.45\\
\hline
XGBClassifier&   	92.5&      91.93&  	92.5&     	92.32\\
\hline
LightGBM&   	92.5&     91.81&  	92.5&   	91.98\\
\hline
\end{tabular}
\label{tab1}
\end{center}
\end{table}

\begin{figure}[htbp]
\centerline{\includegraphics[scale=0.3]{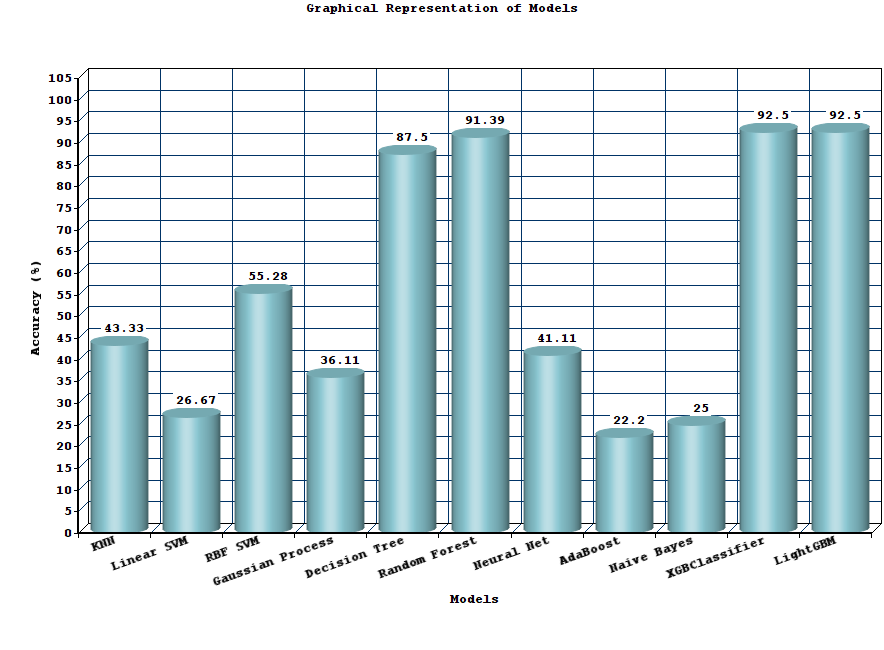}}
\caption{Graphical Representation of Model Accuracy}
\end{figure}

\begin{table}[htbp]
\caption{Best performing classifiers after hyperparameter tuning}
\begin{center}
\begin{tabular}{|c|c|c|c|c|}
\hline
\cline{2-4} 
\textbf{\textit{Model}} & \textbf{\textit{Accuracy}}& \textbf{\textit{Precision}}& \textbf{\textit{Recall}}& \textbf{\textit{F-score}}\\
\hline
Decision Tree&   	92.5(\%)&     92.48(\%)&  	92.5(\%)&     92.28(\%)\\
\hline
Random Forest&   	92.22(\%)&  	91.79(\%)&      92.22(\%)&   	91.95(\%)\\
\hline
\end{tabular}
\label{tab2}
\end{center}
\end{table}
There was a 5 \% jump in accuracy for Decision Tree after hyperparameter tuning and a 0.9 \% increase in accuracy for random forest.

\section{Conclusion}
The purpose of work was to indicate that emotion recognition can be efficiently performed through the amalgamation of low-cost, off the shelf edge devices and advanced machine learning algorithms with low computational resources. We have managed to achieve the same optimally.The results indicate that our method is effectual in emotion recognition. The applications of this study are vast. Mood monitoring is highly relevant to most professional, educational and medical establishments. With the help of proper emotion recognition and classification, mood monitoring can be achieved so as to ensure a mentally healthy population.

\section{Future work}
Future studies would look into the prospect of combining data isolation and personalization for emotion recognition using Federated Learning.Federated learning enables multiple decentralized edge devices to learn a shared predicted model with the training data on device. Mobile devices have become computationally powerful and this can be harnessed to enable on-device training on user specific data, thereby ensuring privacy and personalization.Smart edge device manufacturing companies could tap into the potential of federated learning to enable privacy-oriented emotion monitoring at a large scale by enabling data collection through smart watches/bracelets which are eventually connected to a smart phone wherein the data can be used to train the shared model.The shared model would be trained using the collected data.Then the model is downloaded locally to the user's mobile devices. Smart watches are connected to and controlled through user applications in a mobile phone. These applications can be updated with the sensor data from the smart watch. The data collected can be used to train the model locally with the data specific to the user and the downloaded model.The approach would be data agnostic and user centered.Further,generating real-time data using the readily available smart edge devices to identify the effectiveness of our solution in real-time scenarios over long periods of time while experiencing stimuli would be explored.The applications of this would be vast and profound.

\end{document}